\documentclass[aps,amssymb,showpacs,twocolumn]{revtex4}
\usepackage{amssymb}
\usepackage{graphicx}
\usepackage{color}

\begin{document}

\title{Structure of Compact Stars in Pion Superfluid Phase}
\author{Shijun Mao}
\affiliation{School of Science, Xi'an Jiaotong University, Xi'an 710049, P. R. China}

\begin{abstract}
The gross structure of compact stars composed of pion superfluid quark matter
is investigated in the frame of Nambu-Jona-Lasinio model. Under the Pauli-Villars regularization scheme, the uncertainty of the thermodynamic functions for inhomogeneous states is cured, and  the LOFF state appeared in the hard cutoff scheme is removed from the phase diagram of pion superfluid. Different from the unpaired quark matter
and color superconductor, the strongly coupled pion superfluid is a possible
candidate of compact stars with mass $M\simeq 3M_{\odot }$ and radius $R\simeq 14$ km.
\end{abstract}

\date{\today}
\pacs{26.60.-c, 21.65.Qr, 11.30.Qc}
\maketitle

It is widely accepted that the Quantum Chromodynamics (QCD) phases at high density might be realized in the core of compact stars~\cite{alford1,lattimer,juergen}. For normal quark matter without quark pairings, the equation of state is too soft and thus unable to explain the existence of massive compact stars~\cite{romani,nice}. When diquark pairing is taken into account, the obtained mass-radius relation for compact stars composed of color superconductor~\cite{alford2,rapp} is almost the same as of unpaired quark matter, since the diquark condensate is much less than the corresponding Fermi energy~\cite{shovkovy,stefan,alford3}. However, as estimated from the tree level QCD perturbation theory, the attractive interaction between a quark and an antiquark is stronger than the quark-quark interaction, and thus the equation of state for a pion superfluid~\cite{son,kogut} with large pion condensate will be stiffer in comparison with the color superconductor and may be used to describe the massive compact stars.

Since any QCD phase transition is a nonperturbative phenomenon and its treatment by directly using the QCD itself is still an open question, effective models with QCD symmetries are often used to determine the equation of state of the stellar matter under extreme conditions. One of such models to study the QCD phase structure at finite temperature and density is the Nambu--Jona-Lasinio model (NJL)~\cite{njl1,njl2} at quark level~\cite{vogl,klevansky,volkov,hatsuda,buballa}, which is inspired by the Bardeen-Cooper-Shrieffer (BCS) theory and describes well the quark pairing mechanisms. At zero baryon chemical potential, the quark and antiquark form coherent pairs and condense on a uniform Fermi surface, when the isospin chemical potential is larger than the pion mass $\mu_I > m_\pi$. Inside the pion superfluid phase, there appears a smooth crossover between the BCS condensation of fermions with large and overlapped pairs and the Bose-Einstein condensation (BEC) of molecules with small and distinguished pairs~\cite{sun,he1,mu1,mao,he2}. When the baryon chemical potential is switched on, there appears a Fermi surface mismatch between the quark and antiquark, and the inhomogeneous states, like the Larkin-Ovchinnikov-Fulde-Ferrel state (LOFF)~\cite{ff,lo}, and the gapless Sarma state~\cite{sarma,liu} may enter the phase diagram.

The NJL model with contact interaction between quarks is non-renormalizable, and one requires a regularization scheme to avoid the divergent momentum integrations. A straightforward and widely used scheme is to directly introduce a hard cutoff $\Lambda$ for the quark momentum, which together with the other model parameters can be determined by fitting the quark and meson properties in vacuum. Under such a regularization scheme, one assumes that the temperature and chemical potential of the quark system should be less than the cutoff, $T,\mu <\Lambda$. The NJL model with the cutoff can describe well the phase of chiral symmetry breaking at low temperature and density
and the homogeneous color superconductor and pion superfluid at moderate density. However, when the hard cutoff
is applied to deal with the inhomogeneous LOFF state of relativistic quark systems, unphysical terms occur due
to the lack of the invariance of space translation. It is a nontrival problem to properly renormalize the spurious contribution~\cite{alford4,fukushima1,he3,mu2}.
To avoid the shortcomings arising from the hard cutoff in the study of dense quark matter, we take the
Pauli-Villars regularization scheme in our calculation in the frame of NJL model~\cite{vogl,klevansky,volkov,hatsuda,buballa,florkowski,nickel} where the quark momentum runs formally from zero to infinity.

The two-flavor NJL model at quark level is defined through the Lagrangian density
\begin{equation}
{\cal L} = \bar{\psi}\left(i\gamma^{\mu}\partial_{\mu}-m_0+\gamma_0
\hat{\mu}\right)\psi
+G\left[\left(\bar{\psi}\psi\right)^2+\left(\bar{\psi}i\gamma_5{\bf
\tau}\psi\right)^2 \right]
\label{njl}
\end{equation}
with scalar and pseudoscalar interactions corresponding to $\sigma$
and ${\bf \pi}$ excitations, where $\hat{\mu}
=diag\left(\mu_u,\mu_d\right)=diag\left(\mu_B/3+\mu_I/2,\mu_B/3-\mu_I/2\right)$
is the quark chemical potential matrix with $\mu_u$ and $\mu_d$
being the $u$- and $d$-quark chemical potentials and $\mu_B$ and
$\mu_I$ the baryon and isospin chemical potentials. At $\mu_I=0$, the Lagrangian density has the symmetry of
$U_B(1)\bigotimes SU_I(2)\bigotimes SU_A(2)$ corresponding to baryon
number symmetry, isospin symmetry and chiral symmetry, respectively.
At $\mu_I\ne 0$, the symmetry $SU_I(2)$ firstly breaks down to global $U_I(1)$ symmetry explicitly at $|\mu_I|<m_\pi$, and then the $U_I(1)$ is spontaneously broken at $|\mu_I|>m_\pi$ and the system enters the pion superfluid phase, where $m_\pi$ is the pion mass in vacuum. At $\mu_B=0$, the Fermi surfaces of $u (d)$ and anti-$d
(u)$ quarks coincide, and hence the $\pi_+ (\pi_-)$ condensation is favored at sufficiently high $\mu_I
>0 (\mu_I <0)$. Finite $\mu_B$ provides a mismatch
between the two Fermi surfaces and may lead to gapless or
inhomogeneous LOFF pion condensation. In this case, we
should consider the competition among homogeneous gapped, homogeneous gapless and
inhomogeneous LOFF states.

Since $\mu_I$ is large in the pion superfluid phase, we neglect the possibility of diquark condensation which is favored at large $\mu_B$ and small $\mu_I$ and consider only chiral condensate
\begin{equation}
\label{chiral}
\sigma = \langle\bar{\psi}\psi\rangle,
\end{equation}
and pion condensate
\begin{equation}
\pi e^{-2i{\bf q}\cdot{\bf x}}= \sqrt 2\langle\bar\psi i\gamma_5\tau_-\psi\rangle =
2\langle\bar d i\gamma_5u\rangle
\end{equation}
at $\mu_I>0$. The phase factor $\theta=2 {\bf q} \cdot {\bf x}$ related
to the condensates $\pi$ indicates the direction of
the $U_I(1)$ symmetry breaking. We recover the
homogeneous superfluid state with ${\bf q}=0$ and
obtain the inhomogeneous LOFF state with $\bf q \neq 0$.
Note that we consider here only the single-plane-wave LOFF state for simplicity.

At mean field level, the inverse quark propagator matrix in the flavor space as a
function of quark momentum ${\bf p}$, pair momentum ${\bf q}$, effective pion condensate $\Delta=-2G\pi$ and dynamical quark mass $m=m_0-2G\sigma$ is derived directly,
\begin{eqnarray}
\label{quark}
&&{\cal S}^{-1}({\bf p},{\bf q},\Delta,m)=\\
&&\left(\begin{array}{cc} \gamma^\mu p_\mu-{\bf \gamma}\cdot{\bf q}+\mu_u\gamma_0-m & -i\gamma_5\Delta\\
-i\gamma_5\Delta & \gamma^\mu p_\mu+{\bf \gamma}\cdot{\bf
q}+\mu_d\gamma_0-m\end{array}\right)\nonumber
\end{eqnarray}
and the thermodynamic potential of the system can be expressed as
\begin{equation}
\label{omega}
\Omega(q,\Delta,m) = {1\over 4G}\left[(m-m_0)^2+\Delta^2\right]-\frac{T}{V}\ln \det {\cal
S}^{-1}.
\end{equation}
Note that the thermodynamic potential is a function of
$q=|{\bf q}|$, which means that the direction of ${\bf q}$ is
spontaneously generated and it does not change any physical quantity.
The gap equations to determine physical quantities $m(T,\mu)$, $\Delta(T,\mu)$ and $q(T,\mu)$ at finite temperature and chemical potential can be derived by minimizing the thermodynamic potential,
\begin{equation}
\label{gap}
\frac{\partial\Omega}{\partial m}=\frac{\partial\Omega}{\partial\Delta}={\partial\Omega\over\partial q}=0.
\end{equation}

Once the thermodynamic potential $\Omega$ is known, the thermodynamic functions such as the pressure $P$, entropy density $s$, charge number densities $n_B$ and $n_I$ and energy density $\epsilon$ can be obtained by the thermodynamical relations,
\begin{eqnarray}
\label{thermal}
P&=&-\Omega,\ \ s=-{\partial \Omega\over \partial T},\ \ n_B=-{\partial\Omega\over \partial\mu_B},\ \ n_I=-{\partial\Omega\over \partial\mu_I},\nonumber \\
\epsilon &=&-P+Ts+\mu_I n_I+\mu_B n_B.
\end{eqnarray}

To solve the gap equations and calculate the thermodynamic functions
numerically, we should first fix the model parameters. In any regularization scheme, the NJL model requires three parameters, the
coupling strength $G$, a regulator $\Lambda$ and the current quark
mass $m_0$. They are fixed by fitting the vacuum
properties of the system, such as the quark condensate density $\langle u\bar u\rangle=\langle d \bar
d\rangle$=(-250MeV)$^3$, pion decay constant $f_\pi=93$ MeV and pion mass $m_{\pi}=134$ MeV~\cite{vogl,klevansky,volkov,hatsuda,buballa}.

One of the widely used regularization scheme is the hard
three-momentum cutoff. The procedure is
straightforward --- a cutoff $p<\Lambda$ is
imposed on all momentum integrals after carrying out the Matsubara summation for
$p_0$. This hard cutoff scheme presents reasonable results for the study of homogeneous chiral, diquark and pion condensations~\cite{vogl,klevansky,volkov,hatsuda,buballa}.
However, when the inhomogeneous LOFF state is introduced, such
regularization causes unphysical effects because it
removes the spatial symmetry of the related quasiparticle
spectra~\cite{he1,alford4,fukushima1,he3,mu2}. For instance, the
thermodynamic potential $\Omega(q,\Delta,m)$ outside the pion superfluid should automatically recover the case of free quark gas,
$\Omega(q,0,m)=\Omega(0,0,m)$. However, under the hard cutoff scheme, one can prove
\begin{eqnarray}
\Omega(q,0,m)-\Omega(0,0,m)&=&q{\partial \Omega \over
\partial q}|_{q=0}+{q^2\over 2}{\partial^2 \Omega \over
\partial q^2}|_{q=0}+... \nonumber\\
&=&-{3q^2\over \pi^2}\int_0^\Lambda {p^2 dp\over \sqrt{p^2+m^2}}+...\nonumber\\
&\neq& 0.
\end{eqnarray}
In this case, the thermodynamical potential does not
have a well defined minimum to determine the inhomogeneous equilibrium state.
In order to calculate the phase structure of QCD system, various substraction procedures to avoid the unphysical
terms are proposed, which correspond physically to a vanishing superfluid density~\cite{he3,fukushima2,anderson}.

To have a uniform regularization for the study of homogeneous and inhomogeneous pion superfluid phases, we take a
Pauli-Villars scheme, where one introduces an
arbitrary number of regulating masses $m_j$ and constants $c_j$ and
choose them in such a way that the divergence can be removed by the cancelation among the subtraction terms.

After diagonalize the quark propagator ${\cal S}$, the thermodynamic potential can be expressed as a summation of quasiparticle contributions,
\begin{equation}
\label{thermo2}
\Omega={1\over 4G}\left[(m-m_0)^2+\Delta^2\right]-2N_c \int {d^3
{\bf p}\over (2\pi)^3}\sum_{i=1}^4 g\left[\omega_i({\bf p})\right],
\end{equation}
where $g$ is the thermodynamic distribution function for fermions $g(x)=x/2+T\ln(1+e^{-x/T})$ and $\omega_i$ are the quasiparticle dispersions,
\begin{eqnarray}
\label{quasi}
\omega_1({\bf p})&=&E_+ + \epsilon_-+\mu_B/3,\nonumber\\
\omega_2({\bf p})&=&E_+ - \epsilon_--\mu_B/3,\nonumber\\
\omega_3({\bf p})&=&E_- - \epsilon_-+\mu_B/3,\nonumber\\
\omega_4({\bf p})&=&E_- + \epsilon_--\mu_B/3
\end{eqnarray}
with the definition
\begin{eqnarray}
\label{e}
 E_{\pm} &=& \sqrt{(\epsilon_+ \pm \mu_I/2)^2+\Delta^2}, \\
 \epsilon_\pm &=& \frac{1}{2}\left(\sqrt{({\bf p}+{\bf q})^2+m^2}\pm \sqrt{({\bf p}-{\bf q})^2+m^2}\right).\nonumber
\end{eqnarray}

Under the Pauli-Villars scheme, the summation over the quasiparticles in the potential is regularized as
\begin{equation}
\label{pv}
\sum_{i=1}^4 g[\omega_i] \to \sum_{i=1}^4\sum_{j=0}^Nc_jg[\omega_{ij}],
\end{equation}
and the quasiparticle dispersions $\omega_i, E_\pm$ and $\epsilon_\pm$ are respectively replaced by the regularized ones $\omega_{ij}, E_{\pm j}$ and $\epsilon_{\pm j}$ by regulating the quark mass $m\to m_j=\sqrt{m^2+a_j\Lambda^2}$. The coefficients $a_j$ and $c_j$ are determined by the constraints
\begin{eqnarray}
&&a_0=0,\ c_0=1,\nonumber\\
&&\sum_{j=0}^N c_j m_j^{2L}=0,\ L=0,1,...N-1.
\label{ac}
\end{eqnarray}

While taking $N=2$ in the Pauli-Villars scheme is sufficient to regularize the
quadratic divergencies in the momentum integrations in (\ref{thermo2}) and obtain finite
results from the gap equations, the thermodynamic potential is still
logarithmically divergent, which can then be subtracted by redefining the ground state in vacuum. To regularize the
potential itself, $N=3$ is required. We have numerically checked the regularization
schemes with $N=2$ and $N=3$ and found that they give almost the same results
for the order parameters and the thermodynamical functions. Therefore, we only present in the following the results with $N=3$.

Compared with the hard three-momentum cutoff, the
Pauli-Villars scheme solves the problem of unphysical terms.
The thermodynamic potential outside the inhomogeneous pion superfluid recovers the case of free quark gas, $\Omega(q,0,m)=\Omega(0,0,m)$, since the
quark momentum runs up to infinity and one can make variable shift in momentum integrals, and the zero superfluid density in normal phase is
automatically satisfied. In this case, the gap equations are well defined to determine the ground state of the system, and one needs no longer any substraction.

\begin{figure}[!htb]
\begin{center}
\includegraphics[width=8cm]{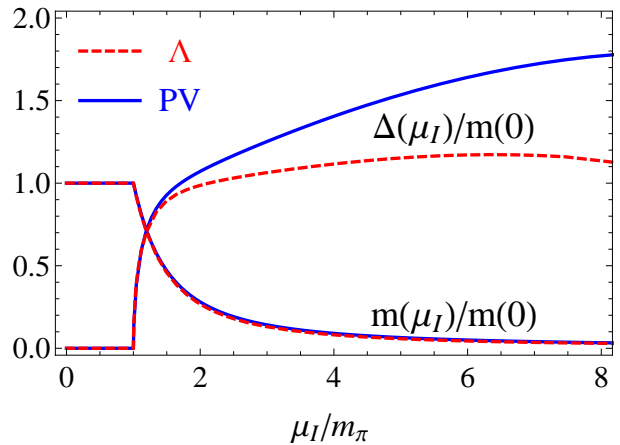}
\caption{(Color online) The scaled pion condensates $\Delta(\mu_I)/m(0)$ and quark mass $m(\mu_I)/m(0)$ as functions of scaled isospin chemical potential $\mu_I/m_\pi$ at $T=\mu_B=0$ in Pauli-Villars (PV, solid lines) and hard three-momentum cutoff ($\Lambda$, dashed lines) schemes. }
\label{fig1}
\end{center}
\end{figure}

We now show the numerical results in the Pauli-Villars scheme (PV) and make comparison with the hard three-momentum cutoff scheme ($\Lambda$). Fig.\ref{fig1} shows the pion condensate $\Delta(\mu_I)$ and quark mass (chiral condensate) $m(\mu_I)$ scaled by the quark mass in vacuum $m(0)$ at $T=\mu_B=0$. Starting from
$\mu_I=m_\pi$, the system enters the
superfluid phase with non-zero pion condensate. As one expected, the chiral condensate which controls the system at low $\mu_I$ is almost independent of the regularization schemes, but the pion condensate which becomes dominant at high $\mu_I$ is sensitive to the scheme we used.
\begin{figure}[!htb]
\begin{center}
\includegraphics[width=8cm]{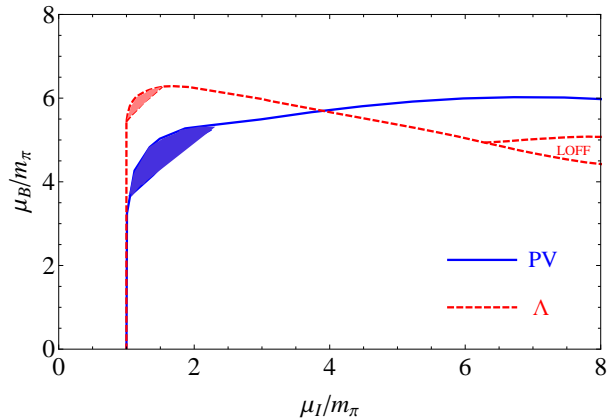}
\caption{(Color online) The pion superfluid phase diagram on the scaled $\mu_I - \mu_B$ plane at $T=0$ in the Pauli-Villars (PV) and hard cutoff ($\Lambda$) schemes. The shaded regions mean the gapless Sarma state, and the LOFF state in the hard cutoff scheme is removed in the Pauli-Villars scheme. }
\label{fig2}
\end{center}
\end{figure}

The pion superfluid phase diagram on the scaled $\mu_I-\mu_B$ plane is shown in
Fig.\ref{fig2}. At $\mu_B=0$ the phase transition
from normal state to pion superfluid happens at $\mu_I=m_\pi$. With increasing $\mu_B$, the mismatch between the Fermi surfaces of $u$ and anti-$d$ quarks reduces the pion condensate. At high enough $\mu_B$, the system comes back to the normal phase. In the
low $\mu_I$ region of the pion superfluid phase, a gapless Sarma state appears near the critical baryon chemical potential under the condition of $\mu_B>3E_{-}|_{min}$ where
the condensate $\Delta$ is lower than in the gapped superfluid state. Since the chemical potentials $\mu_I$ and $\mu_B$ affect the pion superfluid in an opposite way, the critical baryon chemical potential is expected to increase with increasing $\mu_I$, its decrease should be artificial and unphysical. At large $\mu_I$, an inhomogeneous LOFF state between the homogeneous pion superfluid and normal state is predicted in the hard cutoff scheme~\cite{mu2}, see the region surrounded by dashed lines in Fig.\ref{fig2}. This small LOFF region is, however, removed from the phase diagram in the Pauli-Villars scheme. The LOFF state is generally expected to appear in the
weak coupling limit, and the LOFF window becomes more and more narrow when the coupling strength of the matter increases~\cite{son,ff,lo,splittorff,alford5,casalbuoni}. Since the pion superfluid at finite isospin chemical potential is always in a strongly coupled state, indicated by the large condensate~\cite{son} and strong quark potential~\cite{jiang}, the disappearance of the LOFF state in the pion superfluid phase looks reasonable. In fact, the LOFF region under the hard three-momentum cutoff is located
at large $\mu_I$ and $\mu_B$ where the quark chemical potential $\mu_q=\mu_B/3+\mu_I/2$
exceeds already the cutoff $\Lambda$, the calculation in this case becomes quantitatively not reliable. Moreover, the process to subtract the unphysical terms in $\Omega$ in the hard cutoff scheme is not unique and persuasive, which may results in the artificial LOFF state. However, it should be noticed that the NJL model in any regularization scheme is reliable only at low energy scale ($\lesssim 1$ GeV), the possible LOFF state at extremely high isospin chemical potential should be further investigated with other effective methods.

\begin{figure}[!htb]
\begin{center}
\includegraphics[width=8cm]{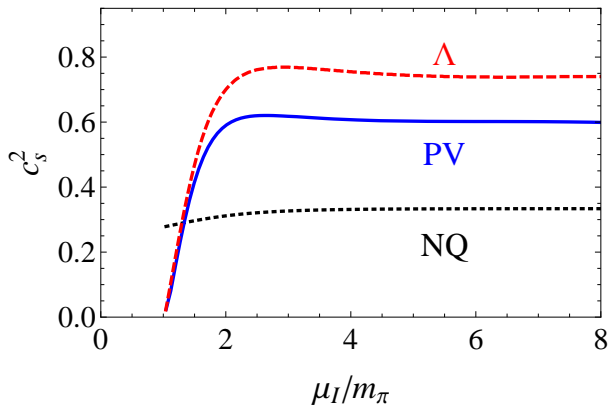}
\caption{(Color online) The sound velocity $c_s^2(\mu_I)$ at $T=\mu_B=0$ in the pion superfluid in Pauli-Villars (PV) and hard cutoff ($\Lambda$) schemes. The dotted line is for the normal quark matter (NQ) at $T=0,\ \mu_B=900 \text{MeV}$ in the Pauli-Villars scheme. }
\label{fig3}
\end{center}
\end{figure}

The coupling strength in a matter can be described by the sound velocity $c_s^2=\partial P/\partial\epsilon$, where $P$ and $\epsilon$ are both monotonic functions of isospin chemical potential $\mu_I$. As shown in Fig.\ref{fig3}, in the pion superfluid it goes up rapidly near the critical point $\mu_I=m_\pi$ and becomes saturated fast with values $c^2_s=0.63$ in Pauli-Villars scheme and
$0.73$ in hard cutoff scheme, and both are much larger than the value $\sim 1/3$ for the
normal quark matter. This indicates clearly that although the pion superfluid is a pairing phenomenon near Fermi surface, the corresponding equation of state is obviously deviated from the normal quark matter due to the strong coupling property at finite isospin chemical potential.

There are many studies on the structure of compact stars in the phase of color superconductivity~\cite{shovkovy,stefan,alford3}. It is found that the color superconductivity does not change the mass-radius relation clearly, since the sound velocity $c_s^2\sim 1/3$ is almost the same as the normal quark matter. Considering the strong coupling shown above at finite isospin chemical potential, we expect a substantial change in the mass and radius of a compact star of pion superfluid.

For a nonrotating and spherically symmetric star, its mass and radius are determined by the Tolman-Oppenheimer-Volkoff (TOV) equations~\cite{t,ov},
\begin{eqnarray}
\label{tov}
&&{d P\over dr}=-{G_N\left(\epsilon+ P\right)\left(M+4\pi r^3 P\right)\over r^2\left(1-2G_N M/r\right)},\nonumber\\
&&{d M\over dr}=4 \pi r^2 \epsilon,
\end{eqnarray}
where $P(r)$ and $\epsilon(r)$ are the pressure and energy density
at radius $r$ inside the star and $M(r)$ is
the total mass contained within a sphere of radius $r$.

\begin{figure}[!htb]
\begin{center}
\includegraphics[width=7.5cm]{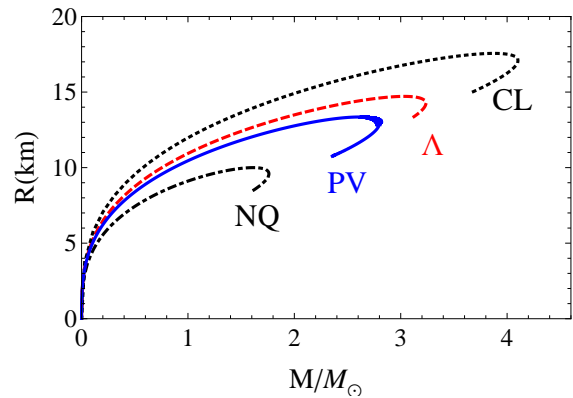}
\caption{The mass-radius relation of compact stars in the state of pion superfluid in Pauli-Villars (PV) and hard cutoff ($\Lambda$) schemes at fixed $T=0$ and $\mu_B=600$ MeV. CL and NQ mean the causal limit with sound velocity $c_s^2=1$ and normal quark matter with $c_s^2=1/3$. }
 \label{fig4}
\end{center}
\end{figure}

Substituting an equation of state $P(\epsilon)$ and giving a fixed central pressure
$P_c=P(r=0)$, one can numerically solving the star mass and
radius by integrating the TOV equations from the center of
the star up to its surface $r=R$ where the pressure reaches its perturbative value $P(R)=B$ with the MIT bag constant $B=75$ MeV fm$^{-3}$~\cite{chodos}. The mass-radius relation for compact stars in the pion superfluid state is shown in Fig.\ref{fig4} at fixed $T=0$ and $\mu_B=600$ MeV. In comparison with the normal quark matter (NQ, dashed-dotted line), the star in the pion superfluid state supports a larger mass and radius, since the equation of state of the pion superfluid with $c_s^2 \gg 1/3$ is much harder. Since the sound velocities in the Pauli-Villars and hard cutoff regularization schemes are similar, the mass-radius relation is not sensitive to the scheme we used. In both cases, the maximum star mass and radius can reach $\sim 3M_{\odot }$ and $\sim 14$ km, while they are only $\sim 1.8M_{\odot }$ and $\sim 10$ km in ideal quark matter. Referring to the
measured data~\cite{juergen,romani,nice}, the pion superfluid is a candidate to explain the massive compact stars like PSR $J1311\text{-}3430$\cite{romani}.
\begin{figure}[!htb]
\begin{center}
\includegraphics[width=7.5cm]{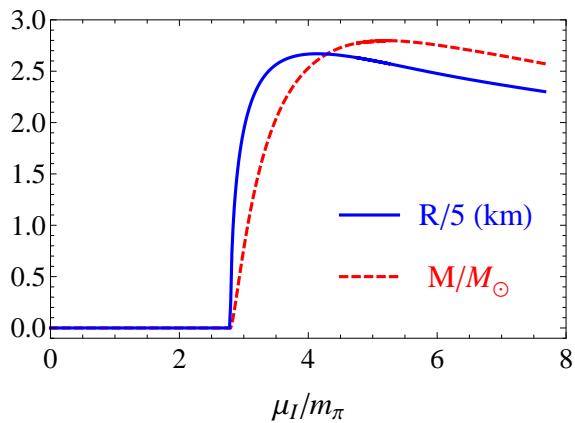}
\caption{The mass and radius $M(\mu_I)$ and $R(\mu_I)$ of compact stars in the state of pion superfluid in Pauli-Villars regularization scheme at fixed $T=0$ and $\mu_B=600$ MeV. }
\label{fig5}
\end{center}
\end{figure}

With increasing isospin chemical potential in the pion superfluid, there exists a crossover from the BEC to BCS states~\cite{sun,he1,mu1,mao,he2}, characterized by the effective chemical potential $\tilde \mu=\mu_I/2-m(\mu_I)$. In the BEC state at low $\mu_I\to m_\pi$, the system can be considered as a weakly coupled pion gas and the maximum pressure $P(0)$ at the center may not reach the condition $P(0)>B$ to form a stable star. In the BCS state at high $\mu_I$, however, the maximum pressure is much larger than the bag constant, $P(0)\gg B$, the quark matter is strongly compressed and it is possible to build up a massive star. Fig.\ref{fig5} shows the star mass and radius as functions of $\mu_I$. A massive star of pion superfluid can exist at high isospin chemical potential.

However, both the radius $R$ and the mass $M$ slowly decrease at high enough $\mu_I$ in Fig.\ref{fig5}, corresponding to the backbends in Fig.\ref{fig4}. This behavior of $R$ and $M$ arises from the TOV equations (\ref{tov}), namely the gravity effect\cite{glend}. At small $\mu_I$, the pressure at the center is small, and it takes only a short distance to reach $P=B$ on the surface. With increasing $\mu_I$, the central pressure increases, and then the distance to reach $P=B$ becomes long. However, when the central pressure is high enough, the gravitational attraction in compact stars becomes strong and leads to a decrease of the radius and mass. Note that the quark stars with $\partial M/\partial P<0$ would
collapse to black holes because of the gravitational instability. In the physical region, both mass and radius increase with increasing pressure or isospin chemical potential.

In summary, we investigated the pion condensation at finite isospin and baryon chemical potentials in the frame of Pauli-Villars regularized NJL model and the structure of compact stars in the state of pion superfluid. Taking the advantage of keeping spatial symmetry in the Pauli-Villars scheme, there is no need to introduce subtraction terms in the model for the study of inhomogeneous pion superfluid, and the LOFF state which appears in the hard cutoff scheme is removed from the model. By solving the TOV equations with the equation of state from the NJL model, the massive stars with mass $M\simeq 3M_{\odot }$ and radius $R\simeq 14$ km can be explained by the strongly interacting pion superfluid with large pion condensate.

\noindent {\bf Acknowledgement:} The work is supported by the NSFC and MOST grant Nos. 11335005, 2013CB922000 and 2014CB845400.

\end{document}